\documentclass[twocolumn,lineno,dvipsnames, usenames]{aastex631}
\usepackage[utf8]{inputenc}
\usepackage{amssymb,amsmath,verbatim,mathtools,needspace,enumitem,etoolbox,graphicx,physics,microtype,ulem, breqn}
\usepackage{xspace}
\usepackage{booktabs}
\usepackage{upgreek}
\usepackage{acronym}
\usepackage{gensymb}

\newcommand{\LIGOlabMIT}{\affiliation{MIT LIGO Laboratory, Massachusetts Institute of Technology, 185 Albany St., Cambridge, MA 02139, USA}}
\newcommand{\MKI}{\affiliation{MIT Kavli Institute for Astrophysics and Space Research, 70 Vassar St., Cambridge, MA 02139, USA}}
\newcommand{\UNLV}{\affiliation{Department of Physics and Astronomy, University of Nevada, Las Vegas, 4505 South Maryland Parkway, Las Vegas, NV 89154, USA}}
\newcommand{\NCfA}{\affiliation{Nevada Center for Astrophysics, University of Nevada, Las Vegas, 4505 South Maryland Parkway, Las Vegas, NV 89154, USA}}

\newcommand{\bilby}{\texttt{Bilby}\xspace}
\newcommand{\bayestar}{\texttt{BAYESTAR}\xspace}

\acrodef{EX}[EX]{\emph{Example}}
\acrodef{bbh}[BBH]{binary black hole}
\acrodef{nsbh}[NSBH]{neutron star--black hole}
\acrodef{bns}[BNS]{binary neutron star}
\acrodef{lvk}[LVK]{LIGO-Virgo-KAGRA}
\acrodef{em}[EM]{electromagnetic}
\acrodef{gw}[GW]{gravitational wave}
\acrodef{agn}[AGN]{active galactic nuclei}
\acrodef{snr}[S/N]{signal-to-noise ratio}
\acrodef{dtd}[DTD]{delay time distribution}
\acrodef{sfr}[SFR]{star formation rate}
\acrodef{cbc}[CBC]{compact-object binary coalescence}
\acrodef{grb}[GRB]{gamma ray burst}
\acrodef{xg}[XG]{next generation}
\acrodef{ce}[CE]{Cosmic Explorer}
\acrodef{et}[ET]{Einstein Telescope}

\date{\today}

\begin{document}

\title{On the use of galaxy catalogs in gravitational-wave parameter estimation}

\correspondingauthor{Geoffrey Mo}
 \email{gmo@mit.edu}
\author[0000-0001-6331-112X]{Geoffrey Mo} \MKI \LIGOlabMIT
\author[0000-0002-7778-3117]{Carl-Johan Haster} \UNLV \NCfA
\author{Erik Katsavounidis} \MKI \LIGOlabMIT

\begin{abstract}
A major challenge in gravitational-wave multi-messenger astrophysics is the imprecise localization of gravitational-wave compact binary mergers. 
We investigate the use of a method to include galaxy catalog information in performing parameter estimation of these events.
We test its effectiveness with the gravitational-wave events GW170817, GW190425, and GW190814, as well as with simulated binary neutron star mergers.
For GW170817, we recover the true host galaxy as the most probable galaxy after a straightforward mass reweighting, with significantly decreased localization area and volume.
On the simulated sample, however, we do not find improvement compared to performing a simple galaxy catalog crossmatch with a regular gravitational wave localization.
Future investigations into sampling methods may yield improvements that increase the viability of this method.
\end{abstract}


\section{Introduction}
\label{sec:intro}
The observation of \acp{gw} by the \ac{lvk} Collaboration has become a regular occurrence.
This has enabled groundbreaking multi-messenger studies of  standard siren cosmology \citep{2017Natur.551...85A, 2023ApJ...949...76A}, neutron star physics \citep{Abbott:2018exr}, r-process nucleosynthesis \citep{2021RvMP...93a5002C}, binary evolution \citep{2018A&A...615A..91B}, and general relativity \citep{2019PhRvL.123a1102A}.
However, much of this science hinges on the precise and accurate localization of \ac{gw} events, which remains a challenging problem.
Traditionally, most \ac{gw} localization is performed using Bayesian methods where an assumption is made that \ac{gw} sources are distributed following a three-dimensional prior distribution that is isotropic over the sky and uniform in comoving volume.
In reality, \ac{gw} events such as \ac{cbc} mergers are generally expected to be hosted in galaxies, which are nonuniform and clustered (see e.g., \citealt{1980lssu.book.....P}), suggesting that folding in existing galaxy catalog information directly into localizations can be beneficial.

Galaxy catalogs have most commonly been used with \ac{gw} data in the context of ``dark siren'' cosmology \citep{Schutz:1986gp, Holz:2005df, 2019ApJ...871L..13F,  2021ApJ...909..218A, Gray:2023wgj}, where \ac{gw} measurements of luminosity distances for mergers without known host galaxies are combined with statistically inferred redshifts from galaxy catalogs to measure the Hubble constant. 
In the case of localization of individual \ac{gw} events, as we discuss here, \citet{2014ApJ...795...43F} performed a proof-of-principle analysis in combining \ac{gw} data with galaxy catalog information.
Their method obtains a joint inferred localization through a post-processing step to regular \ac{gw} parameter estimation, by updating the \ac{gw} posterior information from a galaxy catalog-informed prior.
Other works including \citet{2011CQGra..28h5016W, 2014ApJ...784....8H, 2015ApJ...801L...1B, 2016ApJ...820..136G, 2016MNRAS.462.1085A, Singer:2016eax, 2020MNRAS.495.1841A} have considered the use of galaxy catalogs to optimize searches for \ac{em} counterparts.
These focus on combining the \ac{gw}-only localization with galaxy catalogs, whereas we here consider using \ac{gw} and galaxy catalog data simultaneously to produce a localization.

Here, we present a new method which utilizes information about \ac{cbc} hosts through galaxy catalogs to improve the localization of \ac{gw} events.
The method samples over the known two- or three-dimensional discrete locations of galaxies during the main \ac{gw} parameter estimation analysis, in principle allowing for more precise and efficient estimation of each galaxy's probability of being the true \ac{cbc} host.
We have fully implemented it into \bilby \citep{Ashton:2018jfp, Romero-Shaw:2020owr, KAGRA:2021vkt, LIGOScientific:2021usb}, the standard \ac{gw} parameter estimation tool used by the \ac{lvk}.
We find that our method can significantly reduce localization areas and volumes, especially for nearby \ac{gw} sources ($\lesssim 250$~Mpc) where galaxy catalogs are close to complete.
On GW170817, the only \ac{gw} event with a known host galaxy, our method recovers its host NGC4993, as the most probable galaxy after a simple mass reweighting.
However, when tested on a larger sample of simulated binary neutron star mergers, we find no significant improvement on identifying the true hosts compared to crossmatching the normal localization with a galaxy catalog.

The rest of the paper is laid out as follows. 
In Sec.~\ref{sec:methods}, we describe our method in detail.
We then present its performance on real \ac{gw} observations and on simulated \ac{bns} mergers in Sec.~\ref{sec:results}.
Finally, we contextualize and discuss our method and describe how it may be improved in Sec.~\ref{sec:discussion}.

\section{Method} \label{sec:methods}
Our method requires similar inputs to typical \ac{gw} parameter estimation (\ac{gw} data, calibration uncertainty envelopes, a set of prior distributions etc.), with the addition of a galaxy catalog.
Its outputs are posteriors both with and without the use of the catalog, a ranked list of galaxies in the localization region, and for the 3D case, a combined posterior and skymap.
The method is outlined as follows:
\begin{enumerate}
    \item Perform the regular localization for the \ac{gw} event (i.e., using \ac{gw} data alone), either with a rapid method such as \bayestar \citep{Singer:2015ema, Singer:2016eax, Singer:2016erz} or with full parameter estimation using e.g., \bilby.
    \item Cross-match the output localization with a galaxy catalog, in either a 2D configuration incorporating R.A. and decl., or a 3D configuration, which also includes each galaxy's redshift.
    \item Re-analyze the \ac{gw} data using \bilby with the cross-matched galaxy catalog as an input, assuming equal prior weight for each included galaxy. In addition to the typical parameter estimation outputs, this also results in a posterior over each galaxy in the catalog from step 2.
    \item Produce a ``combined'' localization by sampling over both the usual localization from step 1 and the output from step 3, taking into account the completeness of the galaxy catalog at a given distance.
\end{enumerate}
We describe each step in more detail below.

\subsection{Step 1: Initial localization} \label{subsec:step1}
To reduce the computational cost of sampling over each galaxy in a given catalog, we first compute an initial localization without using information from the galaxy catalog.
This allows us to create a ``subcatalog'' (described in Step 2) which contains only the relevant galaxies, instead of having to sample over the entire galaxy catalog.

The \ac{lvk} has adopted two main algorithms for the localization of \ac{cbc} transients: \bayestar and \bilby.
\bayestar is used in low-latency analyses and runs in $\mathcal{O}$(seconds), whereas \bilby is a more computationally expensive method which performs parameter estimation over the full 17-dimensional \ac{cbc} parameter space.
These algorithms (and \bilby's predecessor \texttt{LALInference}; \citealt{Veitch:2014wba}) have been shown to produce localizations that are largely similar \citep{Singer:2014qca, Berry:2014jja, Farr:2015lna, Ashton:2018jfp, Romero-Shaw:2020owr, Frostig:2021vkt}.
In this work, we use \bilby to produce our initial localizations to enable the comparison and combination (see Sec.~\ref{subsec:step4}) of the resulting skymaps.
If latency is a priority (e.g., for rapid \ac{em} follow-up of \ac{gw} events), \bayestar localizations can be used for this step instead.

\subsection{Step 2: Galaxy catalog cross-match} \label{subsec:step2}
The initial localization produced in Step 1 is then cross-matched with the full galaxy catalog to produce 2-dimensional or 3-dimensional subcatalogs using the \texttt{postprocess.crossmatch} function from \texttt{ligo.skymap} \citep{Singer:2015ema, Singer:2016eax, Singer:2016erz}.
This function takes as inputs a \ac{gw} localization and a list of 2D or 3D coordinates, and returns the enclosed area, volume, and probability at each coordinate.
Our resulting subcatalog contains the galaxies enclosed in the 99\% area (in the 2D case) or volume (3D) of the initial localization.\footnote{This factor is flexible, but we use 99\% in this work to ensure a robust subcatalog.}

Throughout this work, we use the NASA/IPAC Extragalactic Database Local Volume Sample (NED-LVS), galaxy catalog \citep{2023ApJS..268...14C} as our sample catalog.
NED-LVS contains approximately two million galaxies out to a distance of $\sim$1000~Mpc with distance information constructed using data from the NASA/IPAC Extragalactic Database.\footnote{\url{https://ned.ipac.caltech.edu/}}
\citet{2023ApJS..268...14C} provide an estimate of the NED-LVS completeness at each distance, which is important for assessing the validity of our method as well as for combining the resulting posterior with the initial localization, as described in Step 4.

The crucial difference between the 2D and 3D catalogs is the incorporation of redshift (or distance) information. 
Since not all galaxy catalogs include redshifts for all their galaxies, we also allow for the use of 2D catalogs which only contain the R.A. and decl. of the galaxies.

We note that our method is largely agnostic to the choice of catalog: if a user wishes to use e.g., a quasar catalog (such as \citealt{2024ApJ...964...69S}) to test the hypothesis that \ac{bbh} mergers come from \ac{agn} \citep{2019ApJ...884L..50M, 2020PhRvL.124y1102G}, they may do so. 
However, the method performs significantly better with redshift information (see Sec.~\ref{sec:results}), and combining the initial localization with the catalog-informed localization (Step 4) cannot be performed without an estimate of the catalog completeness at each redshift.

\subsection{Step 3: Bilby analysis with galaxy catalog prior} \label{subsec:step3}
Using the subcatalog generated in Step 2, we then perform a re-analysis of the \ac{gw} data utilizing a version of \bilby with an additional prior class.
First, in order to ensure smooth sampling of the subcatalog by \bilby, we sort the galaxies in the subcatalog.
We found that a sorting first by redshift or distance (if using a 3D catalog), then R.A., then decl. was sufficiently smooth for \bilby to successfully sample over the subcatalog.
The subcatalog is passed into \bilby as a prior class, and the index of the subcatalog is sampled assuming a uniform prior and subsequently converted to a two- or three-dimensional coordinate.
We use the NED-LVS specified redshifts, which we then convert to luminosity distances assuming a Planck 2015 cosmology~\citep{Planck:2015fie}.
To ensure complete and robust sampling coverage of the prior subcatalog, we occasionally had to increase the number of nested sampling live points from $1024$ to $\sim 4000$.
The resulting \bilby posterior includes all the usual \ac{gw} parameters, as well as a posterior indicating the support for each galaxy in the subcatalog.\footnote{The posteriors for R.A., decl., and distance are derived from that of the sampled galaxies.}

\subsection{Step 4: Combination of the two posteriors (optional)} \label{subsec:step4}
In the distance regime where galaxy catalogs are close to complete (i.e., within $\sim$250~Mpc; see Figs.~10 and 11 of \citealt{2023ApJS..268...14C}), the output from Step 3 can be sufficient.
However, at larger distances where full catalog completeness can no longer be assumed, we can optionally perform an additional step in combining the initial posterior from Step 1 with the catalog-informed posterior from Step 3.
The combination requires the use of a 3D subcatalog, and an estimate of the galaxy catalog completeness at each redshift.
First, we find the catalog completeness at the redshift of each sample from both the initial and catalog-informed posterior.
Then in the catalog-informed posterior, each sample is kept with probability equal to its completeness; the same is done for the initial posterior but with probability (1 - completeness).
For example, a sample from the catalog-informed posterior with a distance of 10~Mpc will be kept with probability close to 1, since galaxy catalogs are close to complete at such a distance. 
On the other hand, a sample from the initial posterior at 1~Gpc, where catalogs are sparse, will be kept with high probability.
Then, the two lists of kept samples are combined according to a modified weighting $w$ of each run's evidence
\begin{equation}
    w = \frac{V_{\rm 3D}}{V_{\rm init}} \exp(\log\mathcal{Z}_{\rm 3D} - \log\mathcal{Z}_{\rm init}),
\end{equation}
where $V_{\rm 3D}$ is the prior volume of the 3D catalog-informed run, represented by the 99\% volume of the resulting skymap (as detailed in Step 3); $V_{\rm init}$ is the effective prior volume of the initial run, estimated by finding the comoving volume of the shell between the farthest and nearest values in the distance posterior; and $\mathcal{Z}_{\rm 3D}$ and $\mathcal{Z}_{\rm init}$ are the Bayesian evidences from the 3D and initial run respectively.
The combined posterior is then constructed with the kept samples from the 3D run, along with a weight $w$ for the kept samples from the regular run.
A probability skymap can then also be produced from the resulting combined posterior.

\section{Performance}
\label{sec:results}
To evaluate the effectiveness of our method, we tested it on a number of real \ac{gw} events, as well as on simulated \ac{bns} signals.
We use the NED-LVS catalog for all of these tests, generating both 2D (where we ignore the redshift information from NED-LVS) and 3D subcatalogs.

\subsection{GW170817}
\begin{figure}[!htpb]
    \centering
    \includegraphics[width=1\linewidth]{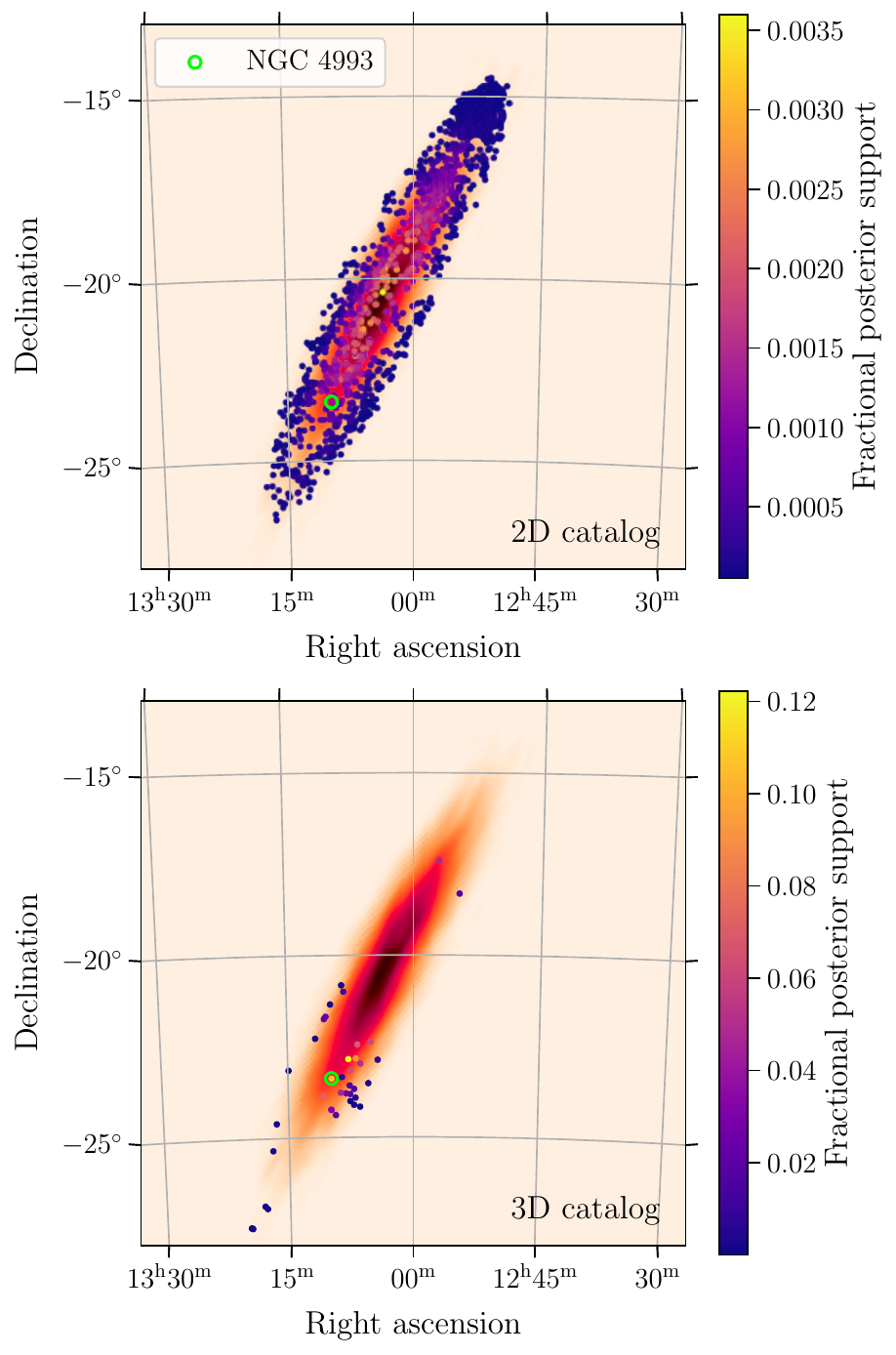}
    \caption{Galaxies in the 2D (top) and 3D (bottom) localization region of GW170817, colored by fractional posterior support as ranked by our method.
    The true host of GW170817, NGC~4993, is highlighted in the green circle.
    Of the 2460 NED-LVS galaxies in the 2D 99\% localization area of GW170817, NGC 4993 was identified as the 305th most likely galaxy by our method.
    In the 3D case, our method ranked NGC 4993 as the second most probable of the 44 galaxies in the 3D 99\% localization volume.
    After performing a stellar mass reweighting, NGC~4993 becomes the most probable host in the 3D case (see text).}
    \label{fig:gw170817_galaxies}
\end{figure}

The canonical \ac{bns} merger GW170817 \citep{LIGOScientific:2017vwq, GBM:2017lvd} is the only \ac{gw} event with a confirmed host galaxy, and is thus a prime test case for our method.
Using the \texttt{IMRPhenomPv2_NRTidalv2} waveform \citep{Hannam:2013oca, Husa:2015iqa, Khan:2015jqa, 2019PhRvD.100d4003D} implemented using a reduced-order quadrature likelihood~\citep{Smith:2016qas, Morisaki:2023kuq}, we first performed the initial parameter estimation step using the noise-subtracted GW170817 \ac{gw} strain \citep{2021SoftX..1300658A} at LIGO~Hanford, LIGO~Livingston \citep{TheLIGOScientific:2014jea}, and Virgo  \citep{TheVirgo:2014hva} (with a low-spin prior and other \bilby configurations consistent with GWTC-2.1~\citep{LIGOScientific:2021usb} and GWTC-3~\citep{KAGRA:2021vkt}).
This resulted in a 90\% localization area of 14.21~deg$^2$ and a 90\% localization volume of 175.9~Mpc$^3$.
Cross-matching the 99\% area and volume localizations to NED-LVS, we find 2460 galaxies in the 2D localization and 44 galaxies in the 3D localization (see Figure~\ref{fig:gw170817_galaxies}).

\begin{figure*}[!htpb]
    \centering
    \includegraphics[width=1\linewidth]{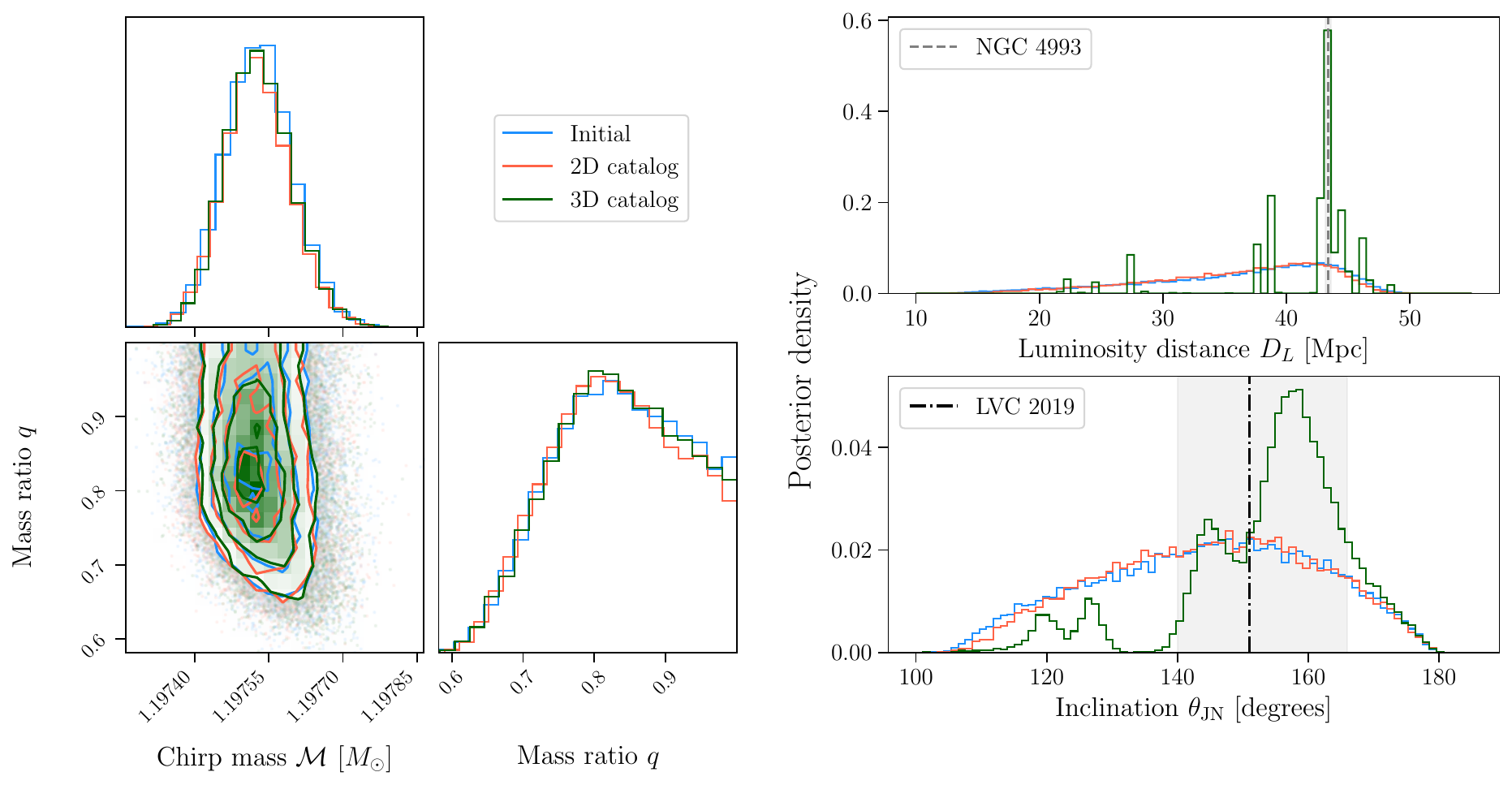}
    \caption{Chirp mass $\mathcal{M}$, mass ratio $q$, luminosity distance $D_L$, and inclination $\theta_{\rm JN}$ posteriors for GW170817.
    The initial (no catalog prior) posteriors are plotted in blue, the 2D catalog posteriors in red, and the 3D catalog posteriors in green.
    The use of the catalog prior does not change the estimation of the intrinsic parameters such as the chirp mass and mass ratio, but improves constraints on extrinsic parameters such as the luminosity distance and inclination angle.
    For comparison with our posteriors, we show the luminosity distance of NGC~4993 as reported in NED-LVS \citep{1991rc3..book.....D, 2023ApJS..268...14C} in the top right panel; in the bottom right panel, we show the median and 90\% confidence interval estimate of the inclination angle of GW170817 as estimated by the LVC in \citet{Abbott:2018wiz} (low-spin prior, \ac{em}-informed result).
    }
    \label{fig:gw170817_posterior}
\end{figure*}

We then performed parameter estimation using our method with both the 2D and 3D subcatalogs.
The fractional posterior support for each galaxy in the 2D and 3D subcatalogs is plotted in Figure~\ref{fig:gw170817_galaxies}, and the posteriors for the chirp mass, mass ratio, luminosity distance, and the inclination $\theta_{\rm JN}$ are shown in Figure~\ref{fig:gw170817_posterior}.
NGC~4993, the true host galaxy, is found as the 305th most likely galaxy out of 2460 in the 2D case, and as the second most probable of 44 galaxies in the 3D localization volume.
Due to the density of the galaxies in the 2D localization, much of the area from the initial localization is sampled, leading to similar posteriors from the initial parameter estimation run and the 2D catalog run, as seen for the luminosity distance and inclination in Figure~\ref{fig:gw170817_posterior}.
For the 3D subcatalog, where there are only 44 possible galaxies, all with redshift information, there are significant improvements in the luminosity distance and inclination constraints.
The luminosity distance of NGC 4993 (using the redshift reported in NED-LVS; \citealt{1991rc3..book.....D}) is strongly preferred by the 3D catalog run.  
The inclination constraint is also stronger, with a large overlap with the inclination angle found in \citet{Abbott:2018wiz} using the \ac{gw} data in addition to the \ac{em}-determined localization and distance.
When restricting the posterior samples to only those corresponding to the top four most probable galaxies, this constraint becomes even more robust.
While estimates of the extrinsic parameters are improved with our method, Figure~\ref{fig:gw170817_posterior}'s left pane shows that intrinsic parameters such as the detector-frame chirp mass $M_c$ and mass ratio $q$ remain unaffected, with essentially identical posteriors across the initial, 2D catalog, and 3D catalog runs.
Since all the support for the luminosity distance falls within 60~Mpc, where the NED-LVS galaxy catalog is close to complete, we do not show the combined posteriors as described in Step 4 (Sec.~\ref{subsec:step4}); they would overlap almost exactly with the 2D- and 3D-catalog posteriors.

A summary of the GW170817 localizations is provided in Table~\ref{tab:gw170817_stats}, showing the dramatic improvement in localization precision and accuracy using our method, especially with a 3D catalog.
In the 2D catalog case, the searched area and volume are larger compared to the initial localization, i.e., the pixel/voxel containing NGC~4993 is less probable compared to others in the skymap.
This is because, as visible in the top panel of Figure~\ref{fig:gw170817_galaxies}, most of the 2460 galaxies in the 2D subcatalog are from two galaxy clusters in the northwest portion of the initial localization.
Since there are fewer galaxies in the ``middle'' of the initial localization, the bulk of the probability is ``pushed'' to those clusters, away from the true host NGC~4993 in the southeast. 
The two galaxy clusters are at approximately 200~Mpc, far outside the initial GW170817 distance posterior, and are thus not included in the 3D-subcatalog (bottom panel of Figure~\ref{fig:gw170817_galaxies}). 
This is an example of the additional power that a 3D catalog provides compared to the 2D case.

\begin{table*}[!htbp]
\centering
  \caption{\label{tab:gw170817_stats} GW170817 localization statistics. The searched area represents the amount of sky that is ranked as more probable than the true location of the event (and similar for searched volume). The Bayes factor reported is that of each 2D- or 3D-catalog run compared to the initial PE run.
  }
    \begin{tabular}{llll}
\toprule
  & Initial PE & 2D-catalog prior & 3D-catalog prior \\
      \midrule
      \midrule
Searched area   [deg$^2$] & 5.21  & 6.86  & 0.00512 \\
Searched volume [Mpc$^3$] & 60.7  & 82.9   & 0.000228 \\
90\% area [deg$^2$]       & 14.2 & 10.6 & 0.898    \\
90\% volume [Mpc$^3$]     & 176 & 119 & 1.54    \\
ln(Bayes factor) & - & 6.83 & 8.95 \\
      \bottomrule
\end{tabular}
\end{table*}

To show the utility of our method for \ac{em} follow-up of \ac{gw} events, we compare our output list of galaxies with those used in the actual search for GW170817's kilonova counterpart.
\citet{Coulter:2017wya} ranked NGC~4993 as the 12th most probable host, while \citet{Arcavi:2017xiz} found it to be the fifth most probable galaxy, and \citet{2017Sci...358.1559K} find it as the third most massive galaxy.
Our list places NGC~4993 second (10.7\% probability), without accounting for stellar mass or luminosity.
When then combining that ranking with the stellar mass as estimated in NED-LVS (simply by multiplying each galaxy's fraction of the total mass in the subcatalog with each galaxy's posterior support fraction as computed by our method), NGC~4993 rises to the top as the most probable host for GW170817.
We show the top ten galaxies in this mass-reweighted list in Table~\ref{tab:ranked_galaxies}.

\subsection{GW190425}
GW190425 \citep{LIGOScientific:2020aai} was the second \ac{bns} merger detected in \acp{gw}.
Because only two \ac{gw} detectors were operating at the time of merger, and because it was located at a much larger distance ($\sim 160$~Mpc), GW190425's localization (8059 deg$^2$ 90\% area) is significantly less precise than that of GW170817.
Using its 99\% localization area and volume, we created subcatalogs with 815,392 and 144,185 galaxies respectively. 
We then applied our method as described above, in the same manner as for GW170817.
Even for a localization region as large as this, we found that our method performed well.

\begin{figure}[!htpb]
    \centering
    \includegraphics[width=1\linewidth]{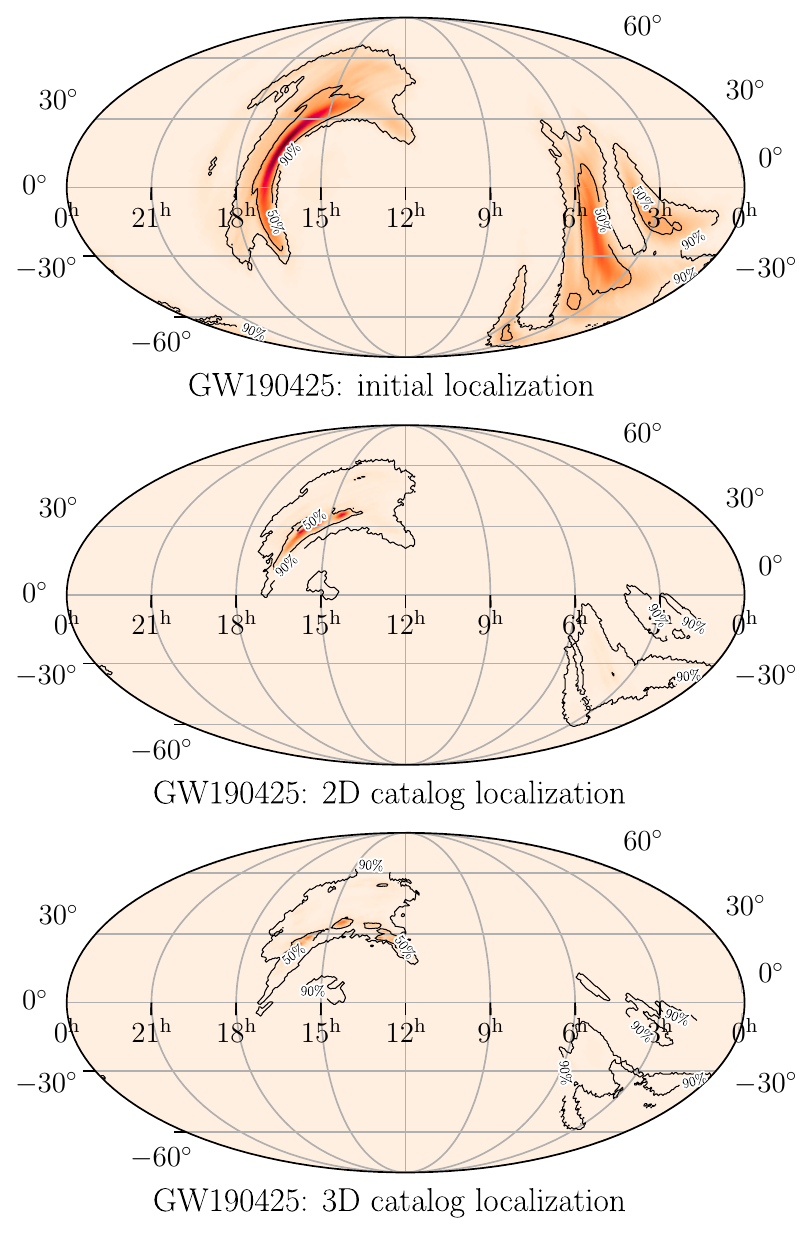}
    \caption{GW190425 initial (top), 2D catalog (center), and 3D (bottom) localizations. Their 90\% areas are 8059, 4086, and 3515 deg$^2$ respectively.}
    \label{fig:gw190425_skymap}
\end{figure}

The initial, 2D, and 3D catalog sky localizations are shown in Figure~
\ref{fig:gw190425_skymap}.
Since almost all the probability falls outside of the Galactic plane and within 250~Mpc, we assume a roughly complete catalog and have not combined the initial posterior with the catalog-informed posteriors (Step 4).

Unlike GW170817, there is no known host for GW190425, so it is impossible to assess the correctness of the localization.
Nevertheless, as with GW170817, we report the ten most highly ranked potential hosts from our 3D analysis, after a the mass reweighting described above, in Table~\ref{tab:ranked_galaxies}.
These ten galaxies comprise 2.9\% of the total reweighted probability.
Since GW190425 is poorly localized, even the most probable galaxies have very little posterior support, indicating that while our method still functions for events with large localizations like GW190425, it is not necessarily as effective as for better-localized events.
The 90\% areas are 8059, 4086, and 3515 deg$^2$ for the initial, 2D catalog, and 3D catalog localizations respectively.

\begin{table*}[!htbp]
\centering
  \caption{\label{tab:ranked_galaxies} The ten most probable galaxies identified from our method, after mass reweighting, using the 3D catalog for GW170817, GW190425, and GW190814. 
  $f_{\rm posterior}$ is the fraction of total posterior support, assuming the galaxy-catalog informed localization alone. 
  $f_{\rm mass}$ is the fraction of the total mass in the 3D subcatalog in a particular galaxy. 
  $f_{\rm posterior} \times f_{\rm mass}$ is the metric by which we rank the galaxies.
  Galaxy names, coordinates, and masses are from the NED-LVS catalog \citep{2023ApJS..268...14C}.
  Note that in this table, we only present galaxies with mass measurements available in NED-LVS.
  }

\begin{tabular}{lllllll}
\toprule
GW170817 & & & & & & \\
\midrule
\midrule
Rank & Galaxy   & R.A. & Decl. & $f_{\rm posterior}$ & $f_{\rm mass}$ & $f_{\rm posterior} \times f_{\rm mass}$ \\
\midrule
1 & NGC 4993 & 197.44875 & $-$23.38389 & $1.06 \times 10^{-1}$ & $5.05 \times 10^{-2}$ & $5.36 \times 10^{-3}$ \\
2 & ESO 575-G023 & 193.65467 & $-$18.30828 & $1.47 \times 10^{-2}$ & $3.03 \times 10^{-1}$ & $4.44 \times 10^{-3}$ \\
3 & IC 4197 & 197.01804 & $-$23.79686 & $2.76 \times 10^{-2}$ & $5.16 \times 10^{-2}$ & $1.42 \times 10^{-3}$ \\
4 & UGCA 331 & 197.69142 & $-$23.86575 & $7.18 \times 10^{-2}$ & $1.81 \times 10^{-2}$ & $1.30 \times 10^{-3}$ \\
5 & ESO 576-G001 & 197.59900 & $-$21.68414 & $2.36 \times 10^{-2}$ & $4.11 \times 10^{-2}$ & $9.73 \times 10^{-4}$ \\
6 & UGCA 327 & 196.93696 & $-$22.85786 & $1.22 \times 10^{-1}$ & $6.41 \times 10^{-3}$ & $7.80 \times 10^{-4}$ \\
7 & ESO 575-G053 & 196.27054 & $-$22.38394 & $6.03 \times 10^{-2}$ & $7.34 \times 10^{-3}$ & $4.43 \times 10^{-4}$ \\
8 & NGC 4968 & 196.77492 & $-$23.67703 & $1.52 \times 10^{-2}$ & $2.86 \times 10^{-2}$ & $4.33 \times 10^{-4}$ \\
9 & ESO 508-G019 & 197.46625 & $-$24.23911 & $3.34 \times 10^{-2}$ & $1.00 \times 10^{-2}$ & $3.35 \times 10^{-4}$ \\
10 & ESO 575-G055 & 196.66631 & $-$22.45606 & $6.98 \times 10^{-2}$ & $3.94 \times 10^{-3}$ & $2.75 \times 10^{-4}$ \\
\bottomrule
\toprule
GW190425 & & & & & & \\
\midrule
\midrule
Rank & Galaxy   & R.A. & Decl. & $f_{\rm posterior}$ & $f_{\rm mass}$ & $f_{\rm posterior} \times f_{\rm mass}$ \\
\midrule
1 & CGCG 109-013 & 246.15412 & 19.50683 & $1.27 \times 10^{-4}$ & $6.58 \times 10^{-4}$ & $8.39 \times 10^{-8}$ \\
2 & NGC 6051 & 241.23612 & 23.93269 & $3.74 \times 10^{-4}$ & $1.53 \times 10^{-4}$ & $5.70 \times 10^{-8}$ \\
3 & UGC 10160 & 240.97685 & 25.01005 & $3.43 \times 10^{-4}$ & $1.41 \times 10^{-4}$ & $4.84 \times 10^{-8}$ \\
4 & CGCG 137-028 & 241.14833 & 25.18993 & $3.12 \times 10^{-4}$ & $1.48 \times 10^{-4}$ & $4.61 \times 10^{-8}$ \\
5 & IC 4569 & 235.20156 & 28.29209 & $3.69 \times 10^{-4}$ & $1.23 \times 10^{-4}$ & $4.56 \times 10^{-8}$ \\
6 & CGCG 166-030 NED01 & 235.14696 & 28.36066 & $4.00 \times 10^{-4}$ & $1.11 \times 10^{-4}$ & $4.45 \times 10^{-8}$ \\
7 & IC 4572 & 235.47580 & 28.13401 & $3.87 \times 10^{-4}$ & $1.12 \times 10^{-4}$ & $4.35 \times 10^{-8}$ \\
8 & IC 0780 & 184.99324 & 25.77170 & $4.00 \times 10^{-4}$ & $1.08 \times 10^{-4}$ & $4.33 \times 10^{-8}$ \\
9 & NGC 3937 & 178.17757 & 20.63131 & $2.59 \times 10^{-4}$ & $1.60 \times 10^{-4}$ & $4.14 \times 10^{-8}$ \\
10 & IC 1219 & 246.11436 & 19.48257 & $2.37 \times 10^{-4}$ & $1.72 \times 10^{-4}$ & $4.09 \times 10^{-8}$ \\
\bottomrule
\toprule
GW190814 & & & & & & \\
\midrule
\midrule
Rank & Galaxy   & R.A. & Decl. & $f_{\rm posterior}$ & $f_{\rm mass}$ & $f_{\rm posterior} \times f_{\rm mass}$ \\
\midrule
1 & ESO 474-G026 & 11.78138 & $-$24.37069 & $1.70 \times 10^{-2}$ & $1.05 \times 10^{-2}$ & $1.79 \times 10^{-4}$ \\
2 & ESO 474-G041 & 13.60150 & $-$25.46408 & $2.07 \times 10^{-2}$ & $5.95 \times 10^{-3}$ & $1.23 \times 10^{-4}$ \\
3 & WISEA J005034.51-233706.7 & 12.64375 & $-$23.61853 & $1.04 \times 10^{-2}$ & $8.34 \times 10^{-3}$ & $8.69 \times 10^{-5}$ \\
4 &IC 1587 & 12.18042 & $-$23.56167 & $7.68 \times 10^{-3}$ & $8.12 \times 10^{-3}$ & $6.24 \times 10^{-5}$ \\
5 & MCG-04-03-016 & 12.25621 & $-$23.81136 & $1.62 \times 10^{-2}$ & $3.16 \times 10^{-3}$ & $5.11 \times 10^{-5}$ \\
6 & WISEA J004610.38-243900.8 & 11.54333 & $-$24.65022 & $1.10 \times 10^{-2}$ & $4.44 \times 10^{-3}$ & $4.88 \times 10^{-5}$ \\
7 & WISEA J005437.69-250401.8 & 13.65696 & $-$25.06717 & $1.76 \times 10^{-2}$ & $2.13 \times 10^{-3}$ & $3.76 \times 10^{-5}$ \\
8 & WISEA J004842.71-234623.2 & 12.17825 & $-$23.77306 & $8.91 \times 10^{-3}$ & $3.81 \times 10^{-3}$ & $3.39 \times 10^{-5}$ \\
9 & WISEA J004854.94-250410.0 & 12.22896 & $-$25.06947 & $2.06 \times 10^{-2}$ & $1.61 \times 10^{-3}$ & $3.31 \times 10^{-5}$ \\
10 & MCG-04-03-017 & 12.31121 & $-$23.85850 & $1.23 \times 10^{-2}$ & $2.67 \times 10^{-3}$ & $3.28 \times 10^{-5}$ \\
\bottomrule
\end{tabular}
\end{table*}

\subsection{GW190814}
We also test our method with GW190814, a highly asymmetric merger of a 23 $M_\odot$ black hole with a 2.6 $M_\odot$ object that is either a black hole or a neutron star.
It was detected by LIGO Hanford, LIGO Livingston, and Virgo, resulting in a 90\% localization of 22~deg$^2$ \citep{2020ApJ...896L..44A}.

\begin{figure*}[!htpb]
    \centering
    \includegraphics[width=1\linewidth]{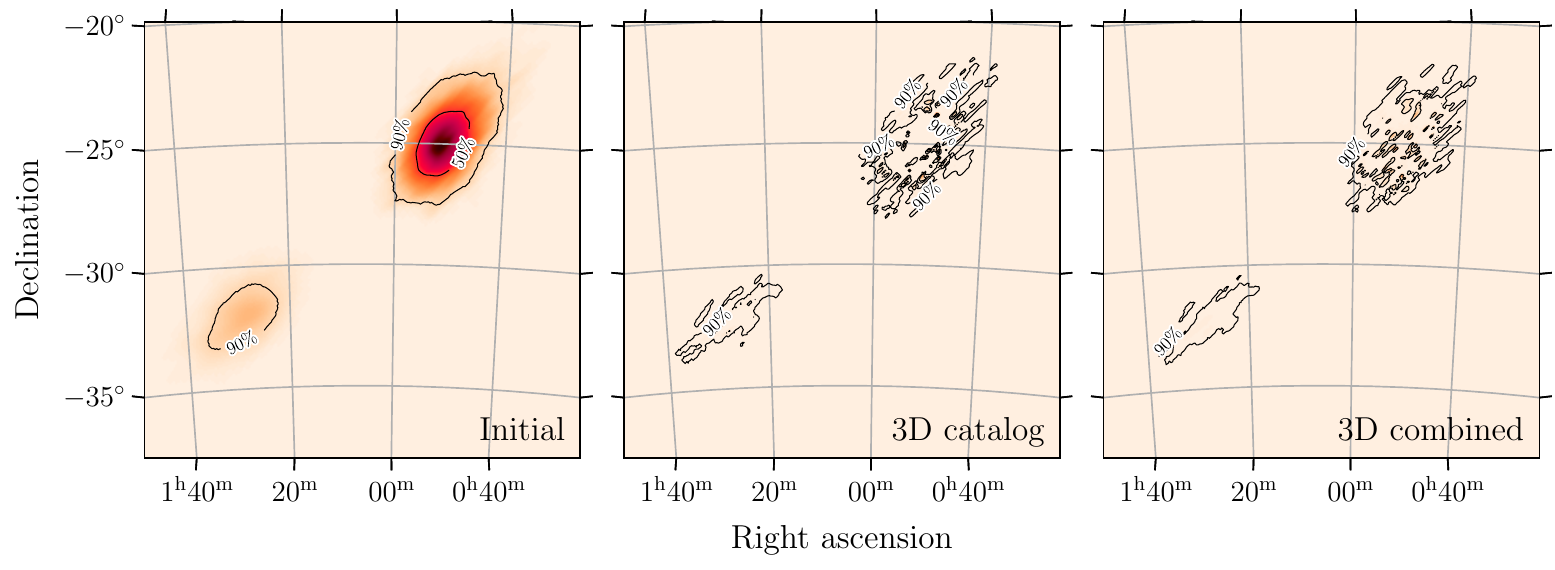}
    \caption{GW190814 initial (left), 3D catalog (center), and combined initial + 3D catalog (right) localizations. Their 90\% areas are 22, 12, and 18 deg$^2$ respectively.}
    \label{fig:gw190814_localizations}
\end{figure*}

As with GW170817 and GW190425, we perform initial, 2D, and 3D localizations for GW190814.
We show the resulting sky maps for the initial, 3D catalog, and combined (see Sec.~\ref{subsec:step4}) initial+3D catalog localizations in Figure~\ref{fig:gw190814_localizations}, to highlight a scenario where the event's larger distance means that the combined localizations have a meaningful effect.
As expected, the combined localization's 90\% area (18~deg$^2$) falls within those of the initial (22~deg$^2$) and 3D catalog (12~deg$^2$) localizations.

\begin{figure}[!htb]
    \centering
    \includegraphics[width=1\linewidth]{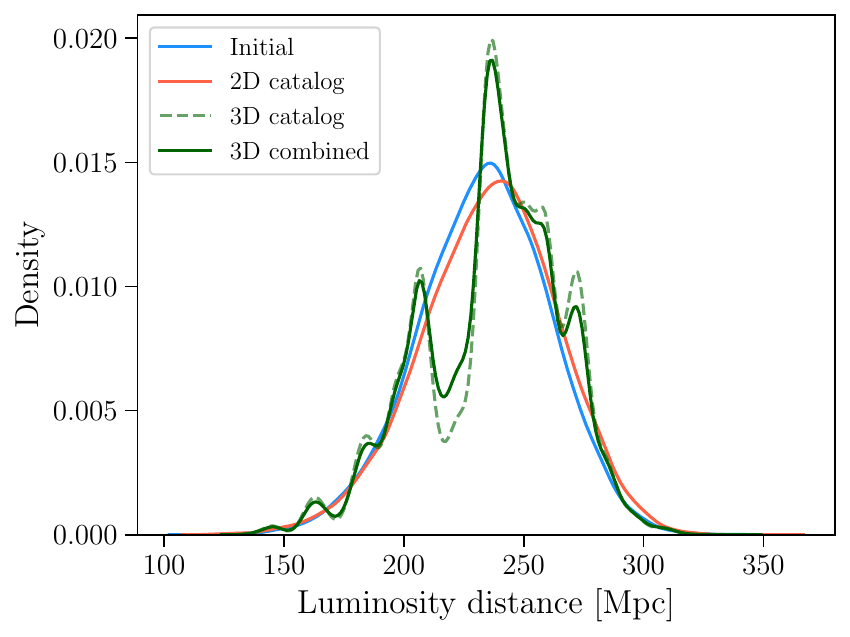}
    \caption{Distance posteriors for GW190814.
    The initial localization is shown in blue, and the 2D catalog localization is shown in orange. For the 3D catalog localizations, we show the catalog-only localizations in the dashed line and the combined (initial + catalog) localization in the solid line.
    Note that the agreement between all posteriors is very good at smaller distances, where NED-LVS is complete, and begins to diverge at larger distances.}
    \label{fig:gw190814_dist}
\end{figure}

We also show the initial, 2D catalog, 3D catalog, and the 3D combined posteriors on the luminosity distance, plotted in Figure~\ref{fig:gw190814_dist}.
Due to the large numbers of galaxies (6138) in the 2D localization, the 2D posterior is not dramatically different from the initial localization.
The 3D posterior, using the 836 galaxies in its subcatalog, is much less uniform, showing two distinct modes around 205 and 240~Mpc.
When combined with the initial localization, the ``peaks'' and ``valleys'' in the posterior become less marked, especially at larger distances, where the NED-LVS completeness falls off.

In addition to the localization, we also produce a ranked list of possible hosts based on the NED-LVS catalog, with the top ten, after a basic mass reweighting, shown in Table~\ref{tab:ranked_galaxies}.

\subsection{Performance on simulated data}

\begin{table*}[!htb]
    \centering
    \caption{\label{tab:bns_sim_params} \ac{bns} simulation parameters.}
    \begin{tabular}{@{}lllllll@{}}
    \toprule
    Parameter             & Distribution \\
    \midrule
    \midrule
    Chirp mass $\mathcal{M}$ & Uniform in components between [1.21, 1.23] $M_\odot$\\
    Mass ratio $q$           & Uniform in components between [0.7, 1.0]\\
    $\chi_1$           & Uniform between [0, 0.05]\\
    $\chi_2$           & Uniform between [0, 0.05]\\
    $\theta_{\rm JN}$           & Uniform in sine\\
    $\psi$           & Uniform between [0, $\pi$]\\
    Phase           & Uniform between [0, $2\pi$]\\
    Luminosity distance $D_L$ & Galaxies with $D_L < 300$~Mpc, $L_{\rm Ks} > 0.1L_{\rm MW}$\\
    R.A. and Decl. & Galaxies with $D_L < 300$~Mpc, $L_{\rm Ks} > 0.1L_{\rm MW}$\\
    \bottomrule
    \end{tabular}
\end{table*}

In addition to the real \ac{gw} events above, we also tested our method on simulated \ac{bns} mergers, where we found mixed results.
Our simulations consisted of 250 \ac{bns} mergers sampled with parameters as described in Table~\ref{tab:bns_sim_params} using the \texttt{IMRPhenomD} waveform model \citep{Husa:2015iqa, Khan:2015jqa, 2023ascl.soft07019P, Morisaki:2023kuq} and a network of LIGO Hanford, LIGO Livingston, and Virgo at O4 design sensitivity\footnote{We use the \texttt{aligo_O4high.txt} and \texttt{avirgo_O4high_NEW.txt} noise curves from \citet{noise_curves}.}, simulated in Gaussian noise.
We place the simulated mergers at the locations of galaxies in the catalog, enforcing only that a host's Ks-band luminosity is greater than 10\% of the Milky Way luminosity to avoid potentially clustered low-mass galaxies\footnote{This reduces the catalog size from 1.9 million to 1.4 million galaxies, a relatively minor effect.}.
We further enforce a network optimal \ac{snr} cut of 8, to avoid very poorly localized events.
131 events pass this \ac{snr} cut, of  which ten events were removed due to problems with sky map generation from posterior samples (for either  the 2D or 3D localizations), resulting in 121 simulated \ac{bns} mergers which we analyze below.\footnote{These ten events did not show obvious trends in their properties, but we acknowledge that this could result in a $\approx 8$\% bias in our cumulative results.}
These simulations are intended to represent a range of realistic \ac{bns} detections for which our method is relevant.

\begin{figure}[!htpb]
    \centering
    \includegraphics[width=1\linewidth]{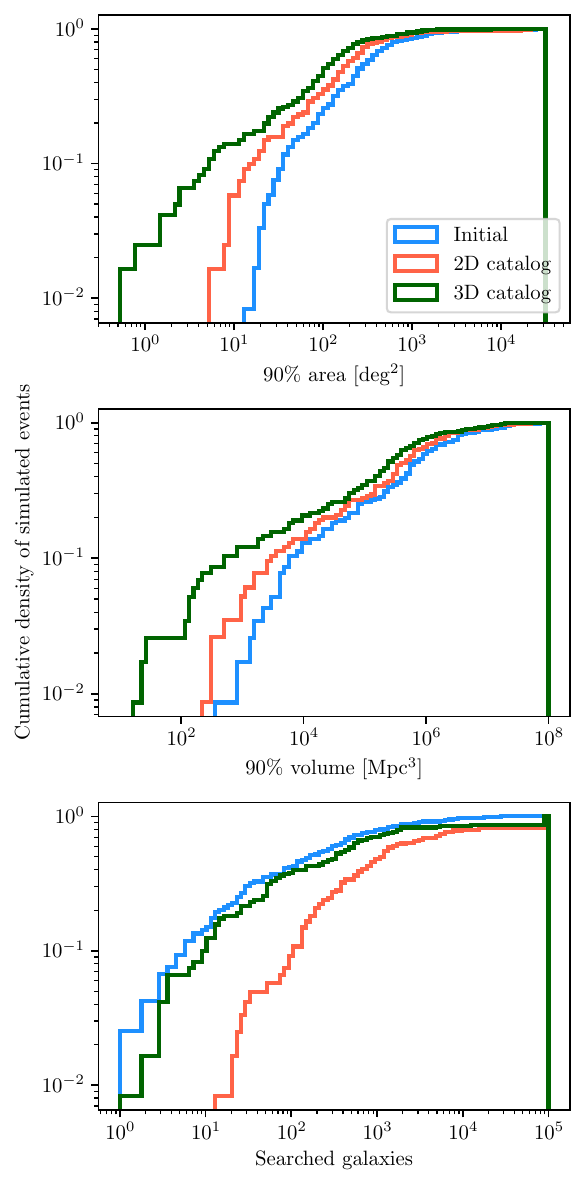}
    \caption{Histograms describing simulated \ac{bns} localization performance.
    \textit{Top:} Cumulative density of the 90\% localization areas of our 121 simulated \ac{bns} mergers. In blue we show the performance of the ``regular'' initial localization with a galaxy catalog crossmatch, and in orange and green we show the performance when using a 2D and 3D galaxy catalog respectively.
    \textit{Middle:} Cumulative density of the 90\% localization volumes of our simulated \ac{bns} mergers.
    \textit{Bottom:} Cumulative density of the number of searched galaxies before arriving at the true simulated host for localizations using 2D (orange) and 3D (green) catalogs. For about 10\% of the simulated \ac{bns} mergers, both with the regular localization crossmatch and when localized using 3D catalogs, the true host is within the top 10 ranked galaxies.}
    \label{fig:sim_histograms}
\end{figure}

Our results are shown as cumulative histograms in Figure~\ref{fig:sim_histograms}.
As with the events detailed above, we find that our method improves significantly on volume and area localizations compared to regular parameter estimation.
In particular, for approximately 10\% of our simulated mergers, the simulated true host is within the top ten ranked galaxies when localized with a 3D catalog.
In some cases, the true host is the top ranked galaxy.

However, we found that the crossmatch of the galaxy catalog with the initial parameter estimation localization (ie., Step 2 described in Sec.~\ref{subsec:step2}), when sorted by the volume searched probability, performs similarly to our method using a 3D catalog\footnote{We do not perform a mass or luminosity weighting in using the 3D catalog, only enforcing that hosts have $L_{\rm Ks} > 0.1L_{\rm MW}$ to match the simulated population.}.
Since this crossmatch is performed with the sky map, which is generated from a kernel density estimate using a subset of the posterior samples, the resulting probabilities at each galaxy location are, in principle, not as precise estimates of the likelihood compared to our method of sampling directly at the discrete position of each galaxy\footnote{See e.g., \citet{2024ApJ...971L...5M} for a demonstration of the possible variance in distance probability density when estimated from posterior samples using different methods. While this likely has a small effect on our results, it is important to keep in mind for counterpart searches using catalog cross-matches from density-estimated posteriors.}.
In practice, our results show that this discrete sampling does not result in significant improvement.
Indeed, for about 13\% of our simulated events, the true host was not sampled by \bilby when localized with the 3D catalog.
This was most prevalent for simulated events at true distances larger than 120~Mpc, where the number of galaxies in the 3D catalog would often be over $10^4$.

The improvement in 90\% area and 90\% volume, while we see no improvement in the distribution of searched galaxies when using a galaxy catalog prior, may seem counterintuitive.
This can be explained by the fact that the use of a galaxy catalog prior necessarily restricts the amount of sky area or volume which can be sampled; thus the resulting area or volume can often be smaller than in the case of the initial localization, even when it is not centered on the true host.
Essentially, the 90\% area and volume are measures of precision, whereas the searched galaxies is a measure of accuracy.

One area where having full posteriors remains advantageous is when considering only the samples corresponding to the most likely few galaxies. 
We found that this results in a more precise and accurate constraint on the inclination angle $\theta_{\rm JN}$ than using all the samples. 
This can also be done in the converse: if we obtain some knowledge about the \ac{gw} inclination angle through the detection of e.g., a poorly localized short \ac{grb} counterpart enforcing that $\theta_{\rm JN} \lesssim 30 \degree$, we can further down-select the list of probable galaxies by only choosing the posterior samples with support at allowed $\theta_{\rm JN}$.

\section{Discussion and conclusion}
\label{sec:discussion}
We have presented a method for using galaxy catalog information in \ac{gw} parameter estimation.
While it showed promise with GW170817, and in principle can provide more precise probabilities for each galaxy being a \ac{cbc}'s host, our simulations concluded that on a population level, it does not outperform simply crossmatching the usual localization with a galaxy catalog.
This shortfall is likely due to the often-large number of galaxies which must be sampled over.
Future work to address different sampling methods, or more sophisticated implementations of galaxy catalog information into parameter estimation tools, may improve the viability of utilizing this important data.
In that light, we offer some thoughts on the usage of galaxy catalogs for this purpose.

In this work, we have used NED-LVS \citep{2023ApJS..268...14C} as an example galaxy catalog, but would like to stress that any galaxy catalog can be used with our method.
For example, large composite catalogs like GLADE \citep{2018MNRAS.479.2374D} and its successor UpGLADE \citep{2024A&A...684A..44B} may extend the utility of our method for mergers at larger distances.
Similarly, an \ac{agn} catalog such as Quaia \citep{2024ApJ...964...69S} could be appropriate for localizing high mass \acp{bbh}, which may reside in \ac{agn} disks \citep{2019ApJ...884L..50M, 2020PhRvL.124y1102G}.
Additionally, since our method does not take into account galaxy parameters such as stellar mass, luminosity, or color, observers may reweight the output list of galaxies freely, e.g, to prioritize galaxies with higher masses or luminosities.
Beyond searches for \ac{em} counterparts where galaxy  parameters are among other considerations (including visibility, airmass, etc.), we note that this reweighting must be done carefully to avoid biases in the context of inferring astrophysical parameters (e.g., $H_0$) from our output galaxies \citep{2024arXiv240514818H, 2024arXiv240507904P}.
Further optimization of host galaxy observing strategies, such as accounting for instrument field-of-view or networks of telescopes, as discussed by \citet{Singer:2016eax} and \citet{Coughlin:2019qkn}, are also possible.

The completeness of the input galaxy catalog is one of the most important factors in using this method.
Existing catalogs such as NED-LVS begin to fall below 70\% K-band completeness at around 350~Mpc ($z \sim 0.075$) \citep{2023ApJS..268...14C}.
Current and future surveys such as SDSS-V \citep{2000AJ....120.1579Y, 2017arXiv171103234K}, DESI \citep{2016arXiv161100036D}, LSST at the Vera Rubin Observatory \citep{2009arXiv0912.0201L, 2019ApJ...873..111I}, Euclid \citep{2011arXiv1110.3193L}, SPHEREx \citep{2014arXiv1412.4872D}, and Roman \citep{2015arXiv150303757S} will push the forefront of survey science, ensuring that galaxy catalogs will continue to deepen.
Future catalogs may enable the use of our method for mergers at higher redshifts.
In a similar vein, powerful upcoming follow-up instruments such as the Vera Rubin Observatory and DSA-2000 \citep{2019BAAS...51g.255H} might be able to produce on-the-fly galaxy catalogs for a given initial localization which may be deeper than existing catalogs \citep{2024MNRAS.533..510M}, enabling an iterative process between \ac{gw} and \ac{em} observers.
Also, if this method can be sped up to $\mathcal{O}$(minute) latencies with further \ac{gw} parameter estimation improvements, rapid catalog-informed localizations could be useful for compute- and storage-limited radio telescope follow-up (e.g., the Long Wavelength Array; \citealt{2018ApJ...864...22A}), where beamforming on the sky is currently an expensive bottleneck.

A specific region where completeness poses a problem is in the Galactic Plane, where extinction complicates observations of galaxies behind the plane at Galactic latitudes $\abs{b} \lesssim 10 \degree$.
One way to mitigate this problem for statistical studies of \acp{gw} localized with our method is by choosing only the ``initial'' posterior for $\abs{b} \lesssim 10 \degree$ in the combination step (Step 4 described in Sec.~\ref{subsec:step4}).
For specific \ac{gw} events where electromagnetic follow-up is the main goal, the impact of the reduction in completeness through the Plane is lessened: if deep IR surveys used to construct catalogs are unable to resolve galaxies behind the plane, limitations on observing resources mean that follow-up campaigns on specific events are also unlikely to be able to detect those hosts.

A caveat that may be addressed in future work is in the use of point estimates for galaxy R.A., decl., and redshift from the galaxy catalog.
In reality, while 2D coordinate errors are typically small compared to telescope fields-of-view, redshift uncertanties, especially for photometric redshifts, can be significant.
A future improvement to our method may take this into account during sampling by marginalising over the estimated localization uncertanties.

Beyond follow-up searches for \ac{em} counterparts to \ac{gw} events, improved localizations incorporating galaxy catalogs may also be useful for dark siren cosmology \citep{Chen:2017rfc, 2021ApJ...909..218A, DES:2019ccw}, though care must be taken in ensuring consistent cosmological models and in accounting for galaxy redshift uncertainties \citep{Turski:2023lxq}.

\section*{}
We would like to acknowledge useful discussions with Gregg Hallinan, Dave Cook, Sylvia Biscoveanu, Salvatore Vitale, and Leo Singer.
G.~M. acknowledges the support of the National Science Foundation and the LIGO Laboratory.
C.-J,~H. acknowledges the support from NASA Grant 80NSSC23M0104 and the Nevada Center for Astrophysics.
LIGO was constructed by the California Institute of Technology and
Massachusetts Institute of Technology with funding from the National
Science Foundation and operates under cooperative agreement PHY-0757058.
The authors are grateful for computational resources provided by the LIGO Lab and supported by NSF Grants PHY-0757058 and PHY-0823459.

This research has made use of data or software obtained from the Gravitational Wave Open Science Center (gwosc.org), a service of the LIGO Scientific Collaboration, the Virgo Collaboration, and KAGRA. This material is based upon work supported by NSF's LIGO Laboratory which is a major facility fully funded by the National Science Foundation, as well as the Science and Technology Facilities Council (STFC) of the United Kingdom, the Max-Planck-Society (MPS), and the State of Niedersachsen/Germany for support of the construction of Advanced LIGO and construction and operation of the GEO600 detector. Additional support for Advanced LIGO was provided by the Australian Research Council. Virgo is funded, through the European Gravitational Observatory (EGO), by the French Centre National de Recherche Scientifique (CNRS), the Italian Istituto Nazionale di Fisica Nucleare (INFN) and the Dutch Nikhef, with contributions by institutions from Belgium, Germany, Greece, Hungary, Ireland, Japan, Monaco, Poland, Portugal, Spain. KAGRA is supported by Ministry of Education, Culture, Sports, Science and Technology (MEXT), Japan Society for the Promotion of Science (JSPS) in Japan; National Research Foundation (NRF) and Ministry of Science and ICT (MSIT) in Korea; Academia Sinica (AS) and National Science and Technology Council (NSTC) in Taiwan.

This paper carries LIGO document number LIGO-P2400420.

Some of the results in this paper have been derived using the \texttt{healpy} \citep{2019JOSS....4.1298Z} and
\texttt{HEALPix} \citep{2005ApJ...622..759G} packages.

\facilities{LIGO, Virgo, KAGRA}

\software{\texttt{astropy} \citep{astropy:2013, Astropy:2018wqo, Astropy:2022ucr},
         \texttt{bilby} \citep{Ashton:2018jfp, Romero-Shaw:2020owr},
         \texttt{healpy} \citep{2019JOSS....4.1298Z},
         \texttt{HEALPix} \citep{2005ApJ...622..759G},
         \texttt{ligo.skymap} \citep{Singer:2015ema, Singer:2016eax, Singer:2016erz},
         \texttt{matplotlib} \citep{Hunter:2007},
         \texttt{numpy} \citep{harris2020array},
         \texttt{pandas} \citep{reback2020pandas, mckinney2010data},
         \texttt{pesummary} \citep{Hoy:2020vys}
        }

\bibliography{galaxy_pe}{}
\bibliographystyle{aasjournal}

\end{document}